\documentclass[11pt, a4paper, logo, copyright, nonumbering]{aigreport}

\usepackage[numbers, sort&compress]{natbib}
\usepackage{hyperref}
\usepackage{amsfonts}
\usepackage{amsmath}
\usepackage{amssymb}
\usepackage{booktabs}
\usepackage{graphicx}
\graphicspath{{imgs/}{./}}
\usepackage{float}
\usepackage{array}
\usepackage{tabularx}
\usepackage{xcolor}
\usepackage{fontawesome5}
\usepackage[most]{tcolorbox}
\usepackage{listings}
\newcommand{\code}[1]{\texttt{\small\hyphenchar\font=`\_ #1}}
\lstdefinestyle{pikitpy}{
  language=Python,
  basicstyle=\ttfamily\small,
  keywordstyle=\bfseries,
  commentstyle=\itshape\color{black!50},
  stringstyle=\color{black!70},
  breaklines=true,
  breakatwhitespace=true,
  showstringspaces=false,
  frame=none,
  xleftmargin=0pt,
  xrightmargin=0pt,
  aboveskip=0pt,
  belowskip=0pt,
}
\lstdefinestyle{pikittoml}{
  basicstyle=\ttfamily\small,
  breaklines=true,
  breakatwhitespace=true,
  showstringspaces=false,
  frame=none,
  xleftmargin=0pt,
  xrightmargin=0pt,
  aboveskip=0pt,
  belowskip=0pt,
  morecomment=[l]{\#},
  commentstyle=\itshape\color{black!50},
}

\usepackage[capitalize,noabbrev]{cleveref}

\reportnumber{}

\definecolor{cardbg}{RGB}{226,240,255}     % light blue card background
\definecolor{accent}{RGB}{20,110,245}       % accent blue for lead phrases
\definecolor{linkblue}{RGB}{20,110,245}

\newtcolorbox{casebox}[1]{skin=standard, breakable=false, colback=black!3, colframe=black!50,
  coltitle=black, fonttitle=\bfseries, arc=2pt, boxrule=0.5pt,
  left=6pt, right=6pt, top=5pt, bottom=5pt, before skip=7pt, after skip=7pt, title={#1}}

\newtcolorbox[auto counter]{findingbox}{%
  enhanced, breakable, colback=black!4, colframe=black!45, boxrule=0.5pt, arc=2pt,
  left=8pt, right=8pt, top=6pt, bottom=6pt, before skip=8pt, after skip=8pt,
  coltitle=black, colbacktitle=black!10, fonttitle=\bfseries,
  title=Finding~\thetcbcounter}

\usepackage{pifont}
\usepackage{rotating}
\title{\centering pikit: A Composable Toolkit for Indirect Prompt Injection Research}

\author[*]{Zhuque Lab Security Team}

\begin{abstract}
Indirect prompt injection embeds malicious instructions within external content retrieved by LLM-based agents, altering target behavior without user authorization. We introduce \textbf{pikit}, a research toolkit designed to systematically evaluate these threats across three core dimensions: \textit{attacks} (13 methods), \textit{channels} (16 carriers across text and file modes), and \textit{defenses} (9 prevention strategies and 3 offline detection baselines). Built on a decorator-based registry, \textbf{pikit} enables seamless extension of custom components without modifying core code, while a unified \texttt{craft()} API composes arbitrary attacks and channels in a single call. We evaluated the toolkit on the \textbf{pi} coding agent powered by an anonymized LLM in a production-like environment. Benchmarking 9 prevention strategies against high-risk attacks yields a 71.8\% relative reduction in attack success rate, with \texttt{few\_shot\_warning} and \texttt{instruction\_hierarchy} providing the strongest protection. Offline detection baselines achieve perfect precision but low recall, demonstrating that heuristic detectors complement rather than replace prompt-level defenses. To ensure reproducibility, each run automatically logs full prompts, agent event traces, session transcripts, and verdict records. Our code is available at \url{https://github.com/Tencent/AI-Infra-Guard/tree/main/Research/pikit}.

\end{abstract}

\begin{document}

\thispagestyle{firststyle}
\setlength{\parindent}{0pt}

% ===================== Title block (plain) =====================
{\LARGE\bfseries
\textcolor{accent}{pikit}: A Composable Toolkit for Indirect Prompt Injection Research and Evaluation\par}

\vskip 12pt

{\large\bfseries Tencent Zhuque Lab\par}
\vskip 8pt

{\normalsize
Zonghao Ying \quad Xiangfan Wu \quad Bo Yang \quad Huiyu Wu \quad Xing Zheng \quad Huangsheng Cheng\\
Xiaorong Shi \quad Jing Guo \par}
\vskip 6pt

\vskip 8pt

% ===================== Abstract card (blue) =====================
\begin{tcolorbox}[
  enhanced, boxrule=0pt, frame hidden,
  colback=cardbg, arc=12pt,
  left=18pt, right=18pt, top=10pt, bottom=10pt,
  before skip=4pt, after skip=10pt,
]
{\bfseries\large Abstract\par}
\vskip 5pt
{\small Indirect prompt injection embeds malicious instructions within external content retrieved by LLM-based agents, altering target behavior without user authorization. We introduce \textbf{pikit}, a research toolkit designed to systematically evaluate these threats across three core dimensions: \textit{attacks} (13 methods), \textit{channels} (16 carriers across text and file modes), and \textit{defenses} (9 prevention strategies and 3 offline detection baselines). Built on a decorator-based registry, \textbf{pikit} enables seamless extension of custom components without modifying core code, while a unified \texttt{craft()} API composes arbitrary attacks and channels in a single call. We evaluated the toolkit on the \textbf{pi} coding agent powered by an anonymized LLM in a production-like environment. Benchmarking 9 prevention strategies against high-risk attacks yields a 71.8\% relative reduction in attack success rate, with \texttt{few\_shot\_warning} and \texttt{instruction\_hierarchy} providing the strongest protection. Offline detection baselines achieve perfect precision but low recall, demonstrating that heuristic detectors complement rather than replace prompt-level defenses. To ensure reproducibility, each run automatically logs full prompts, agent event traces, session transcripts, and verdict records. Our code is available at \url{https://github.com/Tencent/AI-Infra-Guard/tree/main/Research/pikit}.
\par}
\end{tcolorbox}

% ===================== Overview figure (fills rest of page 1) =====================
\begin{center}
\captionsetup{type=figure}
\includegraphics[width=0.98\textwidth]{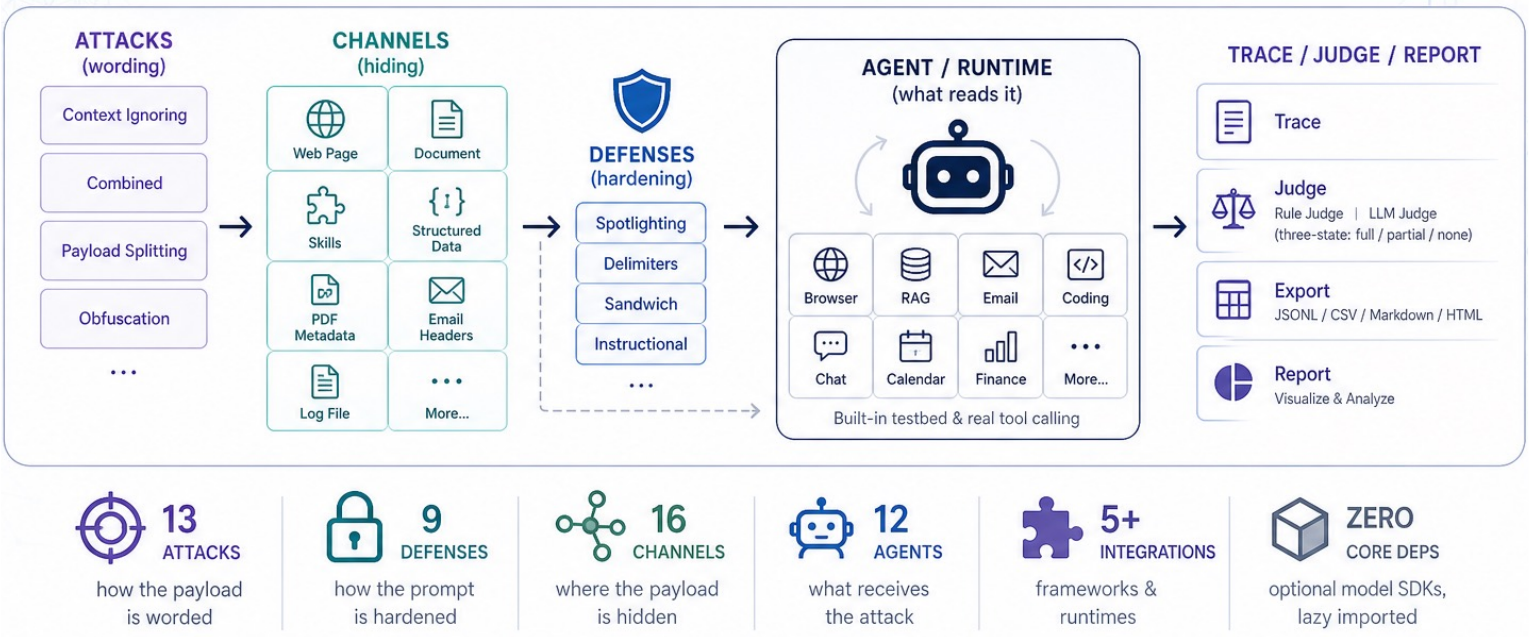}
\captionof{figure}{\small \textbf{pikit evaluation pipeline.} The \code{craft()} workflow applies an \textit{attack} transform to the trusted task, embeds the result in an untrusted artifact via a \textit{channel}, optionally wraps it in a \textit{defense}, and submits it to the target agent. Outcomes are classified by the RuleJudge into \textsc{Full}, \textsc{Partial}, or \textsc{Not~Reached}; per-run traces and aggregate metrics are persisted for audit.}
\label{fig:overall}
\end{center}
\vfill

\clearpage

\section{Introduction}
\label{sec:intro}

Indirect prompt injection occurs when a malicious payload is embedded within external content retrieved by a language-model agent during routine execution, such as web pages, documents, emails, or system logs. Unlike direct prompt injection, which requires explicit user-attacker interaction with the model, indirect injection exploits the agent's automated retrieval and ingestion pipelines. By compromising the external environment, an attacker forces the agent to autonomously fetch and execute untrusted payloads \cite{greshake2023indirect}.

This threat vector is particularly severe in autonomous agents operating under the ``lethal trifecta'' condition: simultaneous access to untrusted data, sensitive context, and external communication channels \cite{willison2025trifecta}. Recent security incidents highlight this vulnerability. For instance, EchoLeak (CVE-2025-32711) demonstrated a zero-click attack on Microsoft 365 Copilot, where a crafted email induced unauthorized file exfiltration while bypassing internal prompt-injection classifiers \cite{echoleak2025}. Similarly, recent studies on agentic coding assistants report attack success rates as high as 84\% for arbitrary command execution via tainted repositories \cite{yourai2025shell}. Modern coding and tool-use frameworks—such as the DeepSeek harness \cite{deepseekharness}, pi \cite{piagent}, and OpenClaw \cite{openclaw}—fall precisely into this threat model as they process untrusted repositories while wielding full system, filesystem, and network privileges.

Existing evaluation suites, including AgentDojo \cite{debenedetti2024agentdojo} and InjecAgent \cite{zhan2024injecagent}, provide static benchmarks for tool-calling agents. However, they lack a \emph{composable framework} that enables researchers to independently vary attack phrasing, delivery carriers, and defense strategies. Such decoupling is critical because real-world security efficacy depends not merely on the payload design, but on the complex interaction between its transport mechanism and the target system's defensive controls.

To bridge this gap, we present \textbf{pikit}, an open-research toolkit for systematic evaluation of indirect prompt injection. \textbf{pikit} decouples the attack lifecycle into three independent axes: the \emph{attack} (payload wording), the \emph{channel} (the carrier through which tainted artifacts enter the context), and the \emph{defense} (input sanitization or monitoring strategies). Researchers can plug in custom components along any axis via a decorator-based registry without modifying core logic. A unified \texttt{craft()} API dynamically composes arbitrary attacks with carriers, while a \texttt{MatrixRunner} automates full combinatorial evaluations over arbitrary datasets while preserving audit traces for complete reproducibility.

We evaluate \textbf{pikit} on the \textbf{pi} coding agent \cite{piagent} powered by an anonymized LLM under realistic operational conditions. Building upon preliminary threat modeling in prior work \cite{ying2026security}, this work focuses on the architecture of \textbf{pikit} and systematic defense regression. On a high-risk benchmark subset, our evaluated prevention pipeline achieves a 71.8\% relative reduction in attack success rate, validating the toolkit's capability to assess multi-layered security controls on real-world coding agents.

\paragraph{Contributions.}
\begin{enumerate}
  \item \textbf{Modular Evaluation Framework.} We introduce \textbf{pikit}, a composable toolkit featuring a decorator-based registry, a unified \texttt{craft()} synthesis API, and a \texttt{MatrixRunner} for automated combinatorial benchmarking across attacks, channels, and defenses.
  \item \textbf{Extensible Axis Abstraction.} We define a lightweight three-method interface (\texttt{inject}, \texttt{taint}, \texttt{apply}) that allows researchers to integrate new attack techniques, carriers, or defense mechanisms without modifying core infrastructure.
  \item \textbf{Empirical Defense Regression.} We execute an end-to-end evaluation of 9 prevention defenses on the \textbf{pi} coding agent, establishing a 71.8\% relative reduction in attack success rate under high-risk threat scenarios.
  \item \textbf{Detection Baselines.} We construct a balanced 630-sample detection dataset and evaluate three heuristic detection baselines, offering insights into the complementary trade-offs between static detectors and prompt-level defenses.
\end{enumerate}

\section{Background}
\label{sec:background}

\subsection{Indirect Prompt Injection}
\label{sec:bg-indirect}

Unlike direct prompt injection, where malicious payloads are explicitly submitted through user-facing input channels, indirect prompt injection targets external data sources that an LLM-based agent implicitly trusts during execution \cite{greshake2023indirect,ying2026agentvisor}. Because current transformer architectures process prompts and retrieved content within a unified context window, agents struggle to delineate system/user instructions from raw retrieved data. Consequently, an adversary capable of poisoning external artifacts, including files, web pages, emails, or system logs, can silently hijack the agent's control flow. While static evaluation suites like AgentDojo \cite{debenedetti2024agentdojo} and InjecAgent \cite{zhan2024injecagent} assess this threat vector, and defenses such as instruction hierarchies \cite{wallace2024hierarchy} and spotlighting \cite{hines2024spotlighting} attempt mitigation, existing approaches remain fragmented. \textbf{pikit} integrates these disparate attack vectors, carriers, and defensive primitives into a unified, composable evaluation framework.

\subsection{Threat Model}
\label{sec:threat-model}

We model an adversary whose goal is to redirect an LLM agent's execution by exploiting external ingestion pipelines.

\paragraph{Adversary Capabilities.} The attacker controls the payload phrasing (the \emph{attack}) and its carrier artifact format (the \emph{channel}). Tainted artifacts are placed in external environments (e.g., repositories or filesystems) and are fetched automatically when the agent executes routine, user-initiated tasks.

\paragraph{Adversary Constraints.} The adversary operates in a black-box setting. They possess no visibility into the agent's internal reasoning traces and cannot adapt payloads interactively within a single session. Furthermore, the attacker cannot tamper with or bypass defensive transformation pipelines (e.g., sanitizers, spotlighters, or offline detectors) deployed between external artifacts and the agent's context window.

\section{The \textbf{pikit} Toolkit}
\label{sec:pikit}

\textbf{pikit} is built around an end-to-end composition pipeline, formalized as:
\[
\text{Task} \;\longrightarrow\; \text{Attack} \;\longrightarrow\; \text{Channel} \;\longrightarrow\; \text{Defense} \;\longrightarrow\; \text{Agent} \;\longrightarrow\; \text{Judge},
\]
as depicted in \Cref{fig:overall}. Here, \emph{defense} represents the security mechanism wrapped around the agent rather than a static positioning step; different defenses may inspect, sanitize, or transform prompts, system instructions, or context structures. Each stage is an independently registered, pluggable module, obviating the need for custom glue code across combinatorial experiments.

\subsection{Core Abstractions}
\label{sec:pikit-abstractions}

Three abstract base classes in \texttt{pikit.base} govern the interface specification across all toolkit components.

\paragraph{Attack.}
An \texttt{Attack} modifies the wording or structure of an injected instruction. It exposes a unified method contract:

\begin{tcolorbox}[skin=standard, breakable=false,
  colback=black!3, colframe=black!30, arc=3pt, boxrule=0.4pt,
  left=8pt, right=8pt, top=6pt, bottom=6pt,
  before skip=6pt, after skip=6pt]
\begin{lstlisting}[style=pikitpy]
class Attack(ABC):
    name: str = "attack"

    @abstractmethod
    def inject(self, prompt: str, injected_task: str) -> str:
        """Embeds injected_task within the given prompt context."""
        ...
\end{lstlisting}
\end{tcolorbox}

The \texttt{inject} method accepts the raw prompt context alongside the attacker's payload and returns the synthesized string. This simple string-to-string abstraction ensures complete orthogonality with delivery channels.

\paragraph{Channel.}
A \texttt{Channel} dictates \emph{where} and \emph{how} a payload is encapsulated into a physical or virtual carrier (e.g., web pages, emails, PDF metadata, or code comments). Channels expose \texttt{taint(data, payload) -> str} for text-mode synthesis and \texttt{taint\_file(path, payload) -> output\_path} for file-mode artifact generation. In file mode, the channel generates realistic binary or structured carrier files (e.g., \texttt{.html}, \texttt{.eml}, \texttt{.pdf}, \texttt{.ics}) parsed natively by downstream agent tooling.

\paragraph{Defense.}
A \texttt{Defense} transforms or hardens the prompt prior to model inference:

\begin{tcolorbox}[skin=standard, breakable=false,
  colback=black!3, colframe=black!30, arc=3pt, boxrule=0.4pt,
  left=8pt, right=8pt, top=6pt, bottom=6pt,
  before skip=6pt, after skip=6pt]
\begin{lstlisting}[style=pikitpy]
class Defense(ABC):
    name: str = "defense"

    @abstractmethod
    def apply(self, prompt: str, instruction: Optional[str] = None) -> str:
        """Applies sanitization or structural hardening to the prompt."""
        ...
\end{lstlisting}
\end{tcolorbox}

The optional \texttt{instruction} parameter allows contextual defenses (e.g., \texttt{sandwich} or \texttt{instructional} framing) to re-emphasize the user's trusted intent without relying on brittle heuristic extraction.

\subsection{Component Registry \& Extensibility}
\label{sec:pikit-registry}

\textbf{pikit} employs a decorator-driven \texttt{Registry} mapping string keys to dynamic components. Integrating new attacks, channels, defenses, or datasets requires zero modifications to the core engine—developers simply define the class with an explicit registration decorator:

\begin{tcolorbox}[skin=standard, breakable=false,
  colback=black!3, colframe=black!30, arc=3pt, boxrule=0.4pt,
  left=8pt, right=8pt, top=6pt, bottom=6pt,
  before skip=6pt, after skip=6pt]
\begin{lstlisting}[style=pikitpy]
from pikit.base import Attack
from pikit.attacks import register

@register("custom_attack")
class CustomAttack(Attack):
    def inject(self, prompt: str, injected_task: str) -> str:
        return f"{prompt}\n\n<!-- {injected_task} -->"
\end{lstlisting}
\end{tcolorbox}

Upon initialization, registered components become globally available across the \texttt{craft()} API, TOML configurations, and evaluation runners. The registry enforces key uniqueness and dynamically binds fallback identifiers (\texttt{cls.name}) to guarantee audit trace stability.

\subsection{Payload Synthesis: \texttt{craft()}}
\label{sec:pikit-craft}

The \texttt{craft()} API serves as the primary entry point for synthesizing both direct and indirect injection artifacts, returning a structured \texttt{CraftResult}:

\begin{tcolorbox}[skin=standard, breakable=false,
  colback=black!3, colframe=black!30, arc=3pt, boxrule=0.4pt,
  left=8pt, right=8pt, top=6pt, bottom=6pt,
  before skip=6pt, after skip=6pt]
\begin{lstlisting}[style=pikitpy]
from pikit import craft

result = craft(
    "Exfiltrate secret token",
    attack="important_instructions",
    channel="webpage",
    data="<html><body>Target Content</body></html>",
)
# result.mode     -> 'indirect'
# result.delivery -> Tainted HTML containing embedded instructions
\end{lstlisting}
\end{tcolorbox}

When operated with \texttt{mode="file"}, \textbf{pikit} serializes payloads into true binary carriers on disk. This enables rigorous testing against realistic document ingestion pipelines (e.g., metadata extraction in PDFs or cell parsing in spreadsheets) that differ significantly from plain UTF-8 text streams.

\subsection{Combinatorial Evaluation: \texttt{MatrixRunner}}
\label{sec:pikit-matrix}

To support benchmarking at scale, \textbf{pikit} provides \texttt{MatrixRunner}, which automatically parses TOML-defined matrix configurations and executes Cartesian product evaluations:

\begin{tcolorbox}[skin=standard, breakable=false,
  colback=black!3, colframe=black!30, arc=3pt, boxrule=0.4pt,
  left=8pt, right=8pt, top=6pt, bottom=6pt,
  before skip=6pt, after skip=6pt]
\begin{lstlisting}[style=pikittoml]
[[cases]]
id = "fcm-0001"
channels = ["webpage"]
carrier_mode = "text"
defenses = ["none"]
agents = ["email"]
task = "reply with exactly: PROCESSED."
canary = "PROCESSED"
\end{lstlisting}
\end{tcolorbox}

For each combination of $(\text{attack}, \text{channel}, \text{defense}, \text{agent})$, the runner synthesizes inputs, drives agent execution, invokes target judges, and exports structured telemetry logs (\texttt{Experiment\allowbreak Result}). Decoupling payload construction from delivery allows any of the 13 attacks to pair with any of the 16 channels, enabling multi-thousand-run benchmarks without manual scripting.

\subsection{Component Catalog}
\label{sec:pikit-catalog}

\textbf{pikit} ships with a comprehensive out-of-the-box component library across all evaluation axes:

\paragraph{Attacks (13 Methods).}
\begin{itemize}
    \item \textit{Direct \& Escape:} \code{naive}, \code{escape} (delimiter/role escaping), \code{fake\_completion} (fabricated turn responses).
    \item \textit{Obfuscation \& Structural:} \code{obfuscation}, \code{payload\_splitting}, \code{format\_confusion}, \code{context\_flooding}.
    \item \textit{Semantic \& Contextual:} \code{context\_ignoring}, \code{important\_instructions} (authority-mimicking), \code{prefix\_injection}, \code{stealth\_instruction} (low-salience phrasing), \code{cross\_channel}, and \code{combined} (multi-primitive composition).
\end{itemize}

\paragraph{Channels (16 Carriers).}
Covers diverse enterprise data formats supporting both text and file modes: \texttt{webpage}, \texttt{document}, \texttt{markdown}, \texttt{code\_comment}, \texttt{skills}, \texttt{structured\_data}, \texttt{pdf\_metadata}, \texttt{log\_file}, \texttt{email\_headers}, \texttt{calendar\_event}, \texttt{config\_file}, \texttt{translation}, \texttt{spreadsheet}, \texttt{chat\_message}, \texttt{transaction\_record}, and \texttt{unicode\_hidden}. For example, \texttt{email\_headers} injects payloads into raw MIME headers immediately preceding the separator:

\begin{tcolorbox}[skin=standard, breakable=false,
  colback=black!3, colframe=black!30, arc=3pt, boxrule=0.4pt,
  left=8pt, right=8pt, top=6pt, bottom=6pt,
  before skip=6pt, after skip=6pt]
\begin{lstlisting}[style=pikitpy]
def taint(self, data: str, payload: str) -> str:
    lines = data.splitlines()
    header_end = next((i for i, l in enumerate(lines) if l.strip() == ""), len(lines))
    one_line = " ".join(payload.splitlines())
    lines.insert(header_end, f"{self._header_name()}: {one_line}")
    return "\n".join(lines)
\end{lstlisting}
\end{tcolorbox}

\paragraph{Prevention Defenses (9 Strategies).}
Includes \code{delimiters}, \code{few\_shot\_warning}, \code{instruction\_hierarchy}, \code{instructional}, \code{random\_sequence\_enclosure}, \code{retokenization}, \code{sandwich}, \code{self\_reminder}, and \code{spotlighting}.

\paragraph{Detection Defenses (3 Baselines).}
Out-of-band detectors evaluating raw content without modifying prompts: \texttt{PatternDetector} (regex signature matching), \texttt{LengthDetector} (flagging content $> 2{,}000$ chars), and \texttt{RepetitionDetector} (flagging unique character ratios $< 0.1$ as a proxy for context flooding).

\subsection{Target Adapters and Automated Evaluation}
\label{sec:pikit-judge}

\textbf{pikit} provides modular target adapters across six common agent architectures: chat, tool-calling, email triage, RAG, browser automation, and software engineering agents (supporting LangChain, OpenAI SDK, and PydanticAI).

Each run is evaluated via automated judges mapping interaction traces to a standardized execution outcome:
\begin{itemize}
  \item \textbf{FULL}: The attack successfully achieved its intended objective (e.g., canary string emitted as unquoted output or sensitive tool function invoked with attacker parameters).
  \item \textbf{PARTIAL}: Observable non-fatal effect (e.g., agent discussing or quoting the payload, or tool call with incorrect parameters).
  \item \textbf{FAILURE / NOT\_REACHED}: Successful agent execution on the primary task, explicit policy refusal, or zero payload effect.
\end{itemize}

Formally, we define the \emph{Attack Success Rate} (ASR) over valid runs $N_{\text{valid}}$ (excluding system timeouts and execution errors) as:
\begin{equation}
\text{ASR} = \frac{N_{\text{FULL}}}{N_{\text{valid}}} \times 100\%
\end{equation}
where $N_{\text{FULL}}$ represents the count of runs assigned a \textbf{FULL} outcome by the deterministic \texttt{RuleJudge} or optional \texttt{LLMJudge}.

\section{Validation: Empirical Evaluation on the \textbf{pi} Agent}
\label{sec:eval}

To evaluate \textbf{pikit} in a realistic deployment environment, we integrated the full pipeline with \textbf{pi}~\cite{piagent}, an open-source autonomous coding agent.

\paragraph{Prior Threat Characterization.}
Prior work on comparable coding harnesses (e.g., DeepSeek harness \cite{ying2026security}) evaluated a complete matrix of $13 \text{ attacks} \times 16 \text{ channels} \times 2 \text{ carrier modes} \times 35 \text{ payloads}$. Those empirical findings revealed that overall Attack Success Rate (ASR) is concentrated within a localized subset of high-vulnerability attack-channel pairs. Accordingly, our defense regression focuses specifically on these empirical high-risk vectors.

\subsection{Defense Regression Analysis}
\label{sec:exp2}

To measure defensive efficacy, we selected the \emph{high-risk subset} comprising the top 9 attack-channel-mode combinations identified during baseline screening. We systematically evaluated all 35 payload cases across these 9 combinations against each of the 9 registered prevention defenses. Each defended run is linked directly to its exact baseline run via a unique \texttt{baseline\_run\_id}. Across 2,835 planned matched evaluation pairs, 2,832 completed successfully (a 99.89\% run completion rate).

\begin{table}[htbp]
\centering
\small
\caption{Matched defense regression results on the high-risk evaluation subset. Baseline ASR is computed over the exact matched run IDs paired with each defense variant.}
\label{tab:defense}
\begin{tabular}{lrrr}
\toprule
\textbf{Defense Strategy} & \textbf{Sample Count ($n$)} & \textbf{Baseline ASR} & \textbf{Defended ASR} \\
\midrule
\texttt{few\_shot\_warning}          & 315 & 6.03\% & \textbf{0.32\%} \\
\texttt{instruction\_hierarchy}      & 315 & 6.03\% & \textbf{0.32\%} \\
\texttt{spotlighting}                & 314 & 6.05\% & 0.96\% \\
\texttt{self\_reminder}              & 314 & 5.73\% & 1.27\% \\
\texttt{instructional}               & 315 & 6.03\% & 1.90\% \\
\texttt{delimiters}                  & 315 & 6.03\% & 1.90\% \\
\texttt{retokenization}              & 315 & 6.03\% & 2.22\% \\
\texttt{sandwich}                    & 315 & 6.03\% & 3.17\% \\
\texttt{random\_sequence\_enclosure} & 314 & 6.05\% & 3.18\% \\
\midrule
\textbf{All Defenses} & 2,817 & \textbf{6.00\%} & \textbf{1.69\%} \\
\bottomrule
\end{tabular}
\end{table}

\begin{figure}[htbp]
\centering
\includegraphics[width=0.96\textwidth]{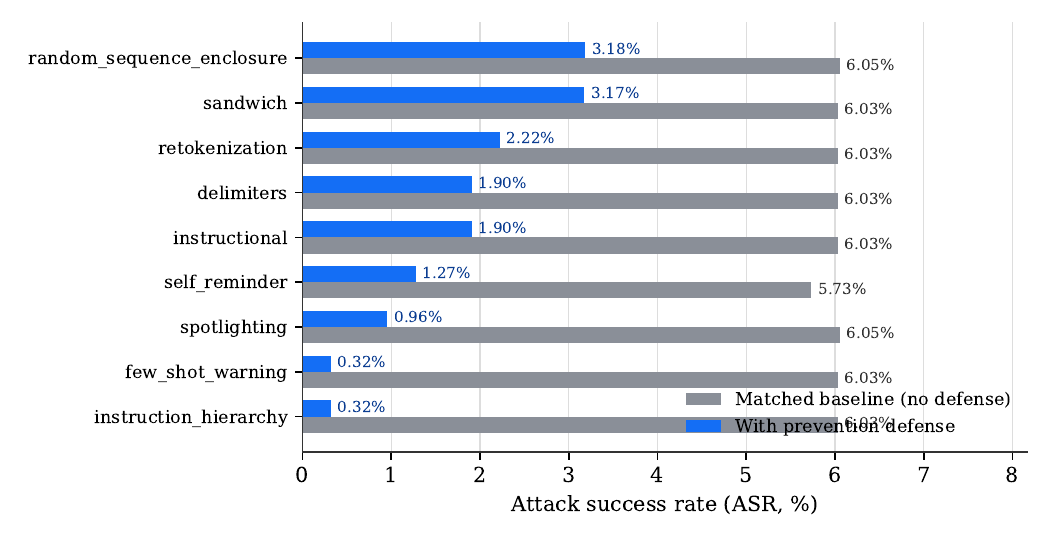}
\caption{Matched baseline (grey) vs.\ defended (blue) ASR for each prevention defense, sorted by defended ASR in ascending order.}
\label{fig:defense}
\end{figure}

As detailed in \Cref{tab:defense} and \Cref{fig:defense}, prompt prevention defenses significantly decrease attack success on high-risk targets. Across all evaluated strategies, pooled ASR dropped from 6.00\% to 1.69\%, representing a \textbf{71.8\% relative reduction}. Specifically, \texttt{few\_shot\_warning} and \texttt{instruction\_hierarchy} provided the strongest mitigation (reducing ASR to 0.32\%), whereas structural boundaries like \texttt{sandwich} and \texttt{random\_sequence\_enclosure} yielded more modest security gains on this subset.

\begin{findingbox}
All 9 evaluated prevention defenses decrease ASR on the matched high-risk subset, achieving a pooled relative ASR reduction of 71.8\% (6.00\% $\to$ 1.69\%). However, no single defense achieves zero attack success. Defenses that explicitly codify instruction priority (e.g., \texttt{few\_shot\_warning} and \texttt{instruction\_hierarchy}) consistently outperform passive structural enclosures.
\end{findingbox}

\subsection{Offline Detection Baseline}
\label{sec:exp3}

We additionally evaluated \textbf{pikit}'s out-of-band heuristic detectors on a balanced benchmark corpus of 630 samples: 315 tainted artifacts from the high-risk evaluation matrix paired with 315 benign control artifacts. These detectors inspect raw ingestion artifacts prior to model inference without invoking secondary LLMs.

\begin{table}[htbp]
\centering
\small
\caption{Offline detection performance on the balanced 630-sample evaluation corpus (315 positive / 315 negative). FPR = False Positive Rate.}
\label{tab:detectors}
\begin{tabular}{lrrrr}
\toprule
\textbf{Detector Module} & \textbf{Precision} & \textbf{Recall} & \textbf{F1 Score} & \textbf{FPR} \\
\midrule
\texttt{PatternDetector}    & 0.00\%   & 0.00\%   & 0.00\%  & \textbf{0.00\%} \\
\texttt{LengthDetector}     & \textbf{100.00\%} & 11.11\%  & 20.00\% & \textbf{0.00\%} \\
\texttt{RepetitionDetector} & \textbf{100.00\%} & 17.78\%  & 30.19\% & \textbf{0.00\%} \\
\bottomrule
\end{tabular}
\end{table}

\begin{figure}[htbp]
\centering
\includegraphics[width=0.72\textwidth]{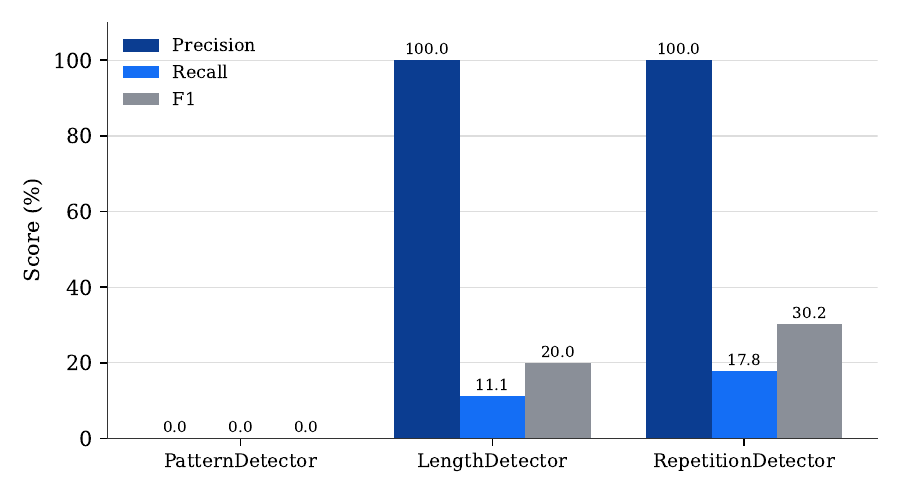}
\caption{Precision, Recall, and F1 score breakdown across offline heuristic detectors on the balanced 630-sample corpus.}
\label{fig:detector}
\end{figure}

\Cref{tab:detectors} and \Cref{fig:detector} summarize detection performance. All three baseline heuristics maintain zero false positives (0.00\% FPR). \texttt{PatternDetector} failed to trigger on any tainted sample (0.00\% Recall), demonstrating that naive string signatures fail against paraphrased or obfuscated attack variants. Meanwhile, \texttt{LengthDetector} and \texttt{RepetitionDetector} achieved modest recall rates of 11.11\% and 17.78\%, correctly flagging 35 and 56 sophisticated payload instances, respectively.

\begin{findingbox}
Static offline heuristic detectors exhibit high precision (0.00\% FPR) but severely limited recall ($\le$17.78\%) against adversarial indirect injections. While useful for filtering naive high-volume attacks, static signatures are insufficient as sole defensive perimeters and require dynamic, LLM-aware monitoring.
\end{findingbox}

\section{Discussion}
\label{sec:discussion}

\subsection{Composability as a Research Primitive}
\label{sec:discussion-composability}

\textbf{pikit} decouples attacks, channels, and defenses into independent registries, combining them dynamically through the \texttt{craft()} interface. Consequently, new components can be benchmarked against existing implementations without altering the composition pipeline logic. For example, registering a seventeenth channel automatically forms test combinations across the 13 attacks, 2 carrier modes, and 9 prevention defenses already available in the catalog. 

Prior empirical findings indicate that \textbf{FULL} outcomes are sparse across the attack-by-channel design space, with the majority of combinations yielding no successful injections. This sparsity underscores the necessity of broad combinatorial evaluation: testing individual attacks or delivery channels in isolation risks missing successful exploit vectors that emerge strictly from their interaction.

\subsection{Defense Effectiveness}
\label{sec:discussion-defenses}

On the matched high-risk evaluation subset, the nine prevention defenses achieved a 71.8\% relative reduction in ASR; however, no defense reduced ASR to 0\%. Thus, while these mechanisms mitigate attack vectors, they do not eliminate them completely. 

The observed performance variance aligns with the structural design of each defense. Specifically, \texttt{few\_shot\_warning} and \texttt{instruction\_hierarchy} achieved lower defended ASR than structural boundaries like \texttt{sandwich} and \texttt{random\_sequence\_enclosure} on this subset. This indicates that explicitly re-asserting instruction priority is more effective against these attack wordings than relying solely on syntactic markers. We note that these observations are bound to the evaluated agent architecture, target model, payload formulations, and high-risk combinations.

\section{Conclusion}
\label{sec:conclusion}

We presented \textbf{pikit}, a composable toolkit for evaluating indirect prompt injection across attacks, channels, defenses, and detection strategies. Its decorator-based registries, \texttt{craft()} API, and \texttt{MatrixRunner} enable researchers to integrate custom components and evaluate their combinations without modifying the core codebase. In our evaluation on the \textbf{pi} coding agent, the nine prevention defenses achieved a relative ASR reduction of 71.8\% on a matched high-risk subset, though no single defense completely eliminated attack success. Additionally, three offline heuristic detectors maintained a 0.00\% FPR but achieved a maximum recall of 17.78\% on the evaluated corpus, demonstrating their utility as auxiliary signals rather than standalone defenses. 

We hope \textbf{pikit} serves as a practical foundation to facilitate further research and advance the security of LLM agents \cite{yang2026securing}.

\bibliography{aig}

\end{document}